\documentclass[11pt,a4paper]{article}
\usepackage[hyperref]{emnlp2018}
\usepackage{times}
\usepackage{latexsym}

\usepackage{url}
\usepackage{graphicx}
\usepackage{fixltx2e}

\usepackage{booktabs}
\usepackage{makecell}

\aclfinalcopy 

\newcommand{\Revised}[2]{\textcolor{orange}{#2}}		

\title{UKP-Athene: Multi-Sentence Textual Entailment for Claim Verification} 

\author{Andreas Hanselowski$^{\dagger}$$^*$, Hao Zhang$^*$, Zile Li$^*$, Daniil Sorokin$^\dagger$$^*$,\\ 
\textbf{Benjamin Schiller$^*$,
Claudia Schulz$^\dagger$$^*$, Iryna Gurevych$^\dagger$$^*$}  \\[2mm]
   $^\dagger$Research Training Group AIPHES \\
   Computer Science Department, Technische Universit\"at Darmstadt\\
   \url{https://www.aiphes.tu-darmstadt.de} \\[2mm]
   $^*$Ubiquitous Knowledge Processing Lab (UKP-TUDA)\\
   Computer Science Department, Technische Universit\"at Darmstadt\\
   \url{https://www.ukp.tu-darmstadt.de/} }

\date{}

\begin{document}
\maketitle

 \begin{abstract}
The Fact Extraction and VERification (FEVER) shared task was launched to support
the development of systems able to verify claims by extracting supporting or refuting facts from raw text. 
The shared task organizers provide a large-scale dataset for the consecutive steps involved in claim verification, 
in particular, document retrieval, fact extraction, and claim classification. 
In this paper, we present our claim verification pipeline approach, 
which, according to the preliminary results, scored third in the shared task, out of 23 competing systems.
For the document retrieval, we implemented a new entity linking approach. In order to be able to rank candidate facts and classify a claim on the basis of several selected facts, we introduce two extensions to the Enhanced LSTM (ESIM).
 \end{abstract}

 

\section{Introduction}

In the past years, the amount of false or misleading content on the Internet has significantly increased.
As a result, information evaluation in terms of fact-checking has become increasingly important
as it allows to verify controversial claims stated on the web. 
However, due to the large number of fake news and hyperpartisan articles published online every day,
manual fact-checking is no longer feasible.  
Thus, researchers as well as corporations are exploring different techniques to automate 
the fact-checking process\footnote{\url{https://fullfact.org/media/uploads/full_fact-the_state_of_automated_factchecking_aug_2016.pdf}}.

In order to advance research in this direction, 
the Fact Extraction and VERification (FEVER) shared task\footnote{\url{http://fever.ai/task.html}} was launched.
The organizers of the FEVER shared task constructed a large-scale dataset~\cite{Thorne18Fever} based on Wikipedia.
This dataset contains 185,445 claims, each of which comes with several evidence sets.
An evidence set 
consists of facts, i.e.~sentences from Wikipedia articles
that jointly support or contradict the claim.
On the basis of (any one of) its evidence sets, each claim is labeled as \emph{Supported}, \emph{Refuted}, 
or \emph{NotEnoughInfo} if no decision about the veracity of the claim can be made.
Supported by the structure of the dataset, the FEVER shared task encompasses three sub-tasks that need to be solved.\\
\textbf{Document retrieval:} Given a claim, find (English) Wikipedia articles containing information about this claim.\\
\textbf{Sentence selection:}
From the retrieved articles, extract facts in the form of sentences
that are relevant for the verification of the claim. \Revised{ "compressed" that 
may form an evidence set and are thus required to determine the veracity of the claim.}{}\\ 
\textbf{Recognizing textual entailment:}
On the basis of the collected sentences (facts), 
predict one of three labels for the claim: \emph{Supported}, \emph{Refuted}, or \emph{NotEnoughInfo}.  
To evaluate the performance of the competing systems, an evaluation metric was devised by the FEVER organizers:
a claim is considered as correctly verified if, in addition to predicting the correct label, a correct evidence set was retrieved.\\
In this paper, we describe the pipeline system that we developed to address the FEVER task.
For document retrieval, we implemented an entity linking approach based on constituency parsing and handcrafted rules.
For sentence selection, we developed a sentence ranking model based on the \emph{Enhanced Sequential Inference Model} (ESIM) \cite{chen2016enhanced}.
We furthermore extended the ESIM for recognizing textual entailment between multiple input sentences and the claim using an attention mechanism.\\
According to the preliminary results of the FEVER shared task,
our systems came third out of 23 competing teams.
The source code for our pipeline, as well as pretrained models, are publicly available\footnote{\url{https://github.com/UKPLab/fever-2018-team-athene}}. 

\section{Background} \label{sec:relwork}

In this section, we present underlying methods that we adopted for the 
development of our system\Revised{for the shared task}{}.

\subsection{Entity linking}
\label{sec:entity-linking}

The document retrieval step requires matching a given claim with the content of a Wikipedia article. A claim frequently features one or multiple \emph{entities} 
that form the main content of the claim.

Furthermore, Wikipedia can be viewed as a knowledge base, where each article describes a particular \emph{entity}, denoted by the article title. Thus, the document retrieval step can be framed as an entity linking problem \cite{Cucerzan2007}. 
That is, identifying entity mentions in the claim and linking them to the Wikipedia articles of this entity. 
The linked Wikipedia articles can then be used as the set of the retrieved documents for the subsequent steps.
%

\subsection{Enhanced Sequential Inference Model}\label{sec:relwork_esim}


Originally developed for the SNLI task \cite{bowman2015large} of determining entailment between two statements, the ESIM (Enhanced Sequential Inference Model) \cite{chen2016enhanced} creates a rich representation of statement-pairs. Since the FEVER task requires the handling of claim-sentence pairs, we use the ESIM as the basis for both sentence selection and textual entailment.
The ESIM solves the entailment problem in three consecutive steps, taking two statements as input.

\noindent\textbf{Input encoding:} 
Using a bidirectional LSTM (BiLSTM) \cite{graves2005framewise}, 
representations of the individual tokens of the two input statements are computed. 

\noindent\textbf{Local inference modeling:}
Each token of one 
statement is used to compute attention weights with respect to each token in the other statement, 
giving rise to an attention weight matrix. 
Then, each token representation is multiplied by all of its attention weights
and weighted pooling is applied to compute a single representation for each token.
This operation gives rise to two new
representations of the two statements.

\noindent\textbf{Inference composition:}
These two statement representations are then fed into two BiLSTMs, 
which again compute sequences of representations for each statement. 
Maximum and average pooling is applied to the two sequences to derive two representations,
which are then concatenated (last hidden state of the ESIM)
and fed into a multilayer perceptron
for the classification of the entailment relation.

\section{Our system for fact extraction and claim verification}\label{sec:system}

In this section, we describe the models that we developed for the three FEVER sub-tasks.

\subsection{Document retrieval}

As explained in Section~\ref{sec:entity-linking}, we propose an \emph{entity linking} approach to the document retrieval sub-task. 
That is, we find entities in the claims that match the titles of Wikipedia articles (documents).
Following the typical entity linking pipeline, we develop a document retrieval component that has three main steps. 

\noindent\textbf{Mention extraction:}
Named entity recognition tools focus only on the main types of entities (Location, Organization, Person). In order to find entities of different categories, such as movie titles, that are numerous in the shared task data set, we employ the constituency parser from AllenNLP \cite{Gardner2017AllenNLP}. After parsing the claim, we consider every \emph{noun phrase} as a potential entity mention. 
However, a movie or a song title may be an adjective or any other type of syntactic phrase. To account for such cases,
we use a heuristic that adds all words in the claim before the main verb as well as the whole claim itself as potential entity mentions. For example, a claim ``\emph{Down With Love is a 2003 comedy film.}'' contains the noun phrases `\emph{a 2003 comedy film}' and `\emph{Love}'. Neither of the noun phrases constitutes an entity mention, but the tokens before the main verb, `\emph{Down With Love}', form an entity.

\noindent\textbf{Candidate article search:}
We use the MediaWiki API\footnote{\url{https://www.mediawiki.org/wiki/API:Main_page}} to search through the titles of all Wikipedia articles for matches with the potential entity mentions found in the claim. The MediaWiki API uses the Wikipedia search engine to find matching articles. The top match is the article whose title has the largest overlap with the query. 
For each entity mention, we store the seven highest-ranked Wikipedia article matches. 

The MediaWiki API uses the online version of Wikipedia and since there are some discrepancies between the 2017 dump used in the shared task and the latest version, we also perform an exact search over all Wikipedia article titles in the dump. We add these results to the set of the retrieved articles.

\noindent\textbf{Candidate filtering:}
The MediaWiki API retrieves articles whose title overlaps with the query. Thus, the results may contain articles with a title longer or shorter than the entity mention used in the query. Similarly to previous work on entity linking \cite{sorokin2018mixing}, we remove results that are longer than the entity mention and do not overlap with the rest of the claim. To check this overlap, we first remove the content in parentheses from the Wikipedia article titles and stem the remaining words in the titles and the claim. Then, we discard a Wikipedia article if its stemmed article title is not completely included in the stemmed claim.

We collect all retrieved Wikipedia articles for all identified entity mentions in the claim after filtering and supply them to the next step in the pipeline.
The evaluation of the document retrieval system on the development data shows the effectiveness of our ad-hoc entity linking approach (see Section \ref{sec:results}).

\subsection{Sentence selection}
In this step, we select candidate sentences as a potential evidence set 
for a claim from the Wikipedia articles retrieved in the previous step.  
This is achieved by extending the ESIM to generate a ranking score on the basis of two input statements, instead of predicting the entailment relation between these two statements.

\begin{figure}[t]
{\centering
\includegraphics[width=0.9\linewidth]{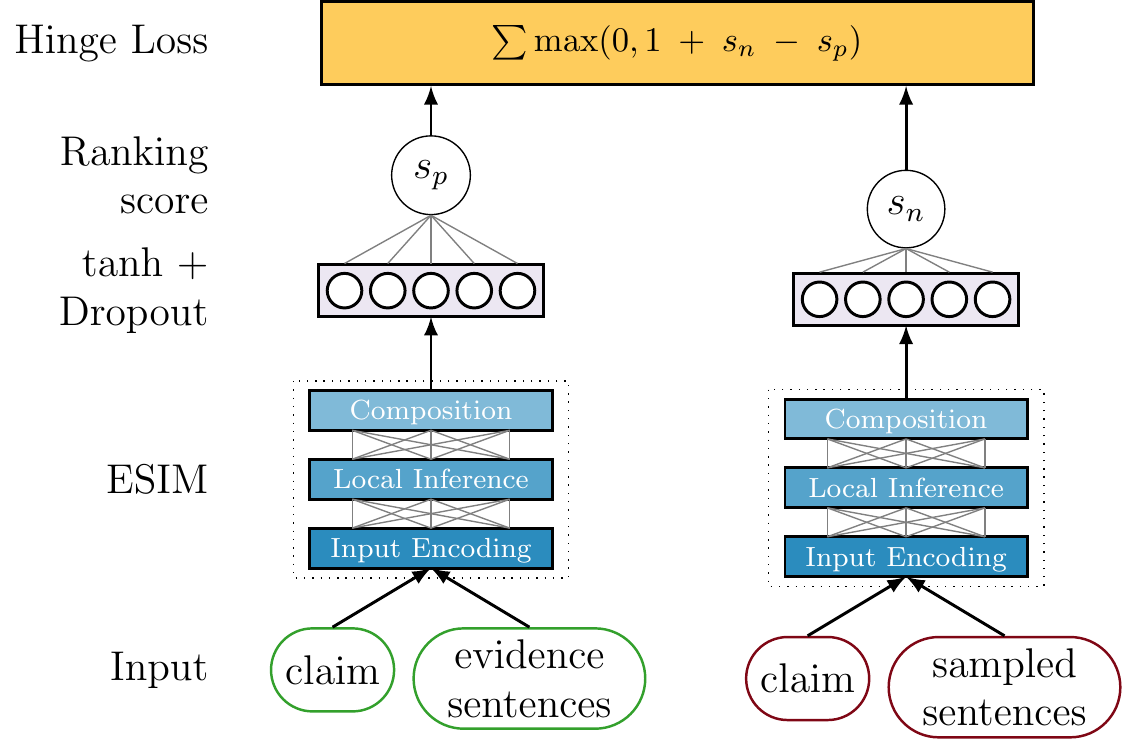}
\caption{Sentence selection model}
\label{fig:senSel}}
\end{figure}

\noindent\textbf{Architecture:}
The modified ESIM takes as input a claim and a sentence. 
To generate the ranking score, the last hidden state of the ESIM (see Section~\ref{sec:relwork_esim}) is fed into 
a hidden layer which is connected to a single neuron for the prediction of
 the ranking score.
As a result, we are able to rank all sentences of the retrieved documents according to the computed ranking scores.
In order to find a potential evidence set, we select the five highest-ranked sentences.

\noindent\textbf{Training:} 
Our adaptation of the ESIM is illustrated in Fig.~\ref{fig:senSel}. 
In the training mode, the ESIM takes as input a claim and the concatenated sentences of an evidence set.
As a loss function, we use a modified hinge loss with negative sampling: 
$\sum max(0,\,1+s_n-s_p)$, where $s_p$ indicates the positive ranking score and $s_n$ the negative ranking score 
for a given claim-sentence pair.
To get $s_p$, we feed the network a claim and the concatenated sentences of one of its ground truth evidence sets. 
To get $s_n$, 
we take all Wikipedia articles from which the ground truth evidence sets of the claim originate, randomly sample five sentences 
(not including the sentences of the ground truth evidence sets for the claim), 
and feed the concatenation of these sentences into the same ESIM. 
With our modified hinge loss function, we then try to maximize the margin between positive and negative samples.

\noindent\textbf{Testing:}
At testing time, we calculate the score between a claim and each sentence  in the retrieved documents.
For this purpose, we deploy an ensemble of ten models with 
different random seeds. 
Then, the mean score of a claim-sentence pair over all ten models of the ensemble is calculated and the scores for all pairs are ranked.
Finally, the five sentences of the five highest-ranked pairs are taken as an output of the model.

\subsection{Recognizing textual entailment}

\begin{figure}[t]
\includegraphics[width=\linewidth]{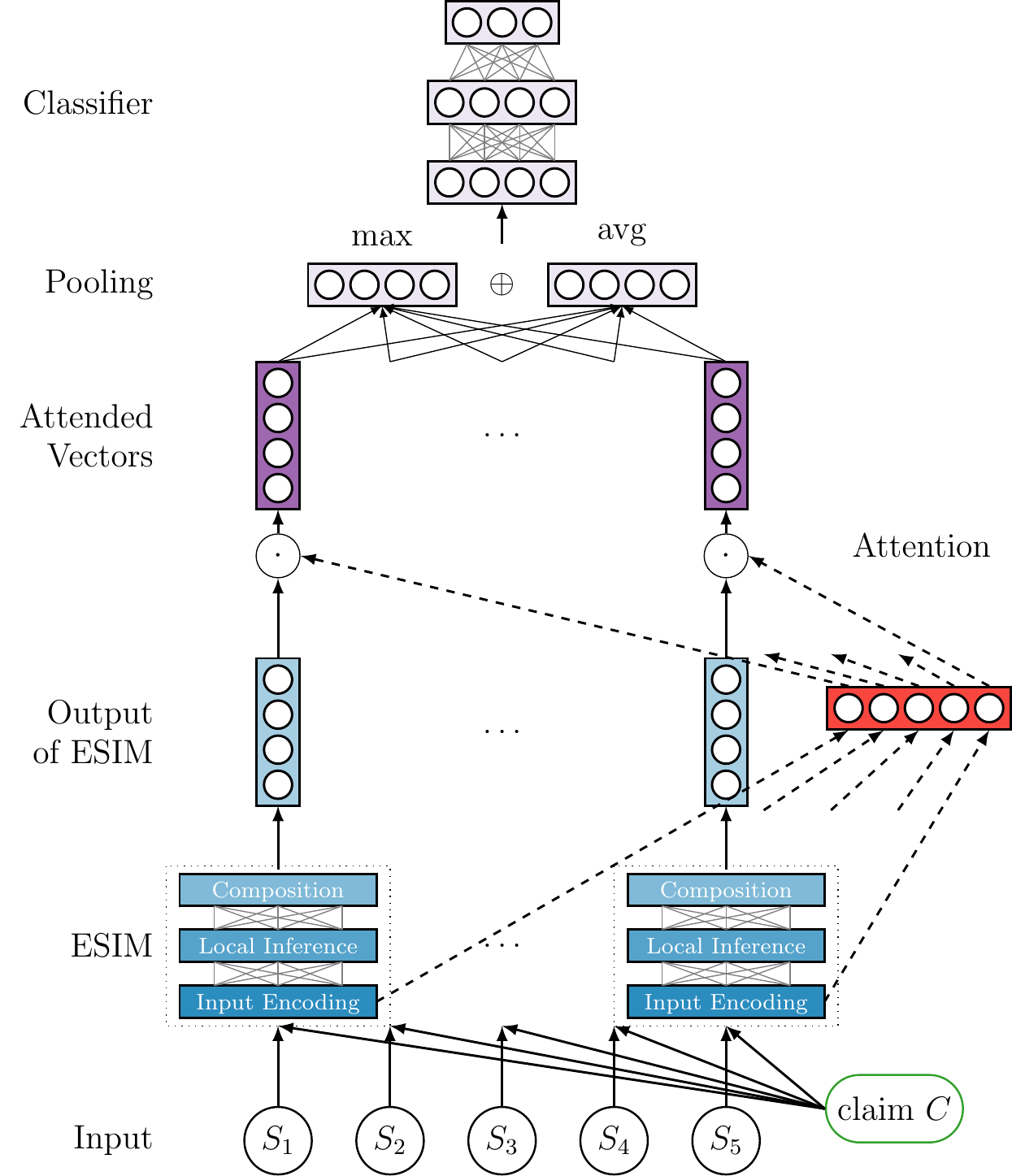}
\centering
\caption{Extended ESIM for recognizing textual entailment}
\label{fig:rte_classifier}
\end{figure}

In order to classify the claim as $Supported$, $Refuted$ or $NotEnoughInfo$,
we use the five sentences retrieved by our sentence selection model described in the previous section. 
For the classification, we propose another extension to the ESIM,
which can predict the entailment relation between \emph{multiple} input sentences and the claim. 
Fig.~\ref{fig:rte_classifier} gives an overview of our extended ESIM for the FEVER textual entailment task.

As word representation for both claim and sentences, we concatenate the Glove \cite{pennington2014glove} 
and FastText \cite{bojanowski2016enriching} embeddings for each word.
Since both types of embeddings are pretrained on Wikipedia, they are particularly suitable for our problem setting. 


To process the five input sentences using 
the ESIM, we combine the claim with each sentence and feed the resulting pairs into the ESIM. 
The last hidden states of the five individual claim-sentence runs of the ESIM are compressed into one vector using attention and pooling operations.\\
The attention is based on a representation of the claim that is independent of the five sentences. This representation is obtained by summing up the \emph{input encodings} of the claim in the five ESIM runs.
In the same way, we derive five sentence representations, one from each of the five runs of the ESIM, which are independent of the claim.
For each claim-sentence pair, the single sentence representation and 
the claim representation are then individually fed through a single layer perceptron. The cosine similarity of these two vectors is then used as an attention weight. 
The five output vectors of all ESIMs are multiplied with their respective attention weights
and we apply average and max pooling on these vectors in order to reduce them to two representations.
Finally, the two representations are concatenated and fed through a 3-layer perceptron to predict one of the three classes
\emph{Supported}, \emph{Refuted} or \emph{NotEnoughInfo}.
The idea behind the described attention mechanism is to allow the model to extract information from the five sentences that is most relevant for the classification of the claim. 


\section{Results}\label{sec:results}

Table~\ref{tb:retrieval_search} shows the performance of our document retrieval and sentence selection system when retrieving different numbers of the highest-ranked Wikipedia articles.
In contrast to the results reported in Table~\ref{tb:results}, here we consider a single model instead of an ensemble. 
The results show that both systems benefit from a larger number of retrieved articles.
\begin{table}[h]
\centering
\begin{tabular}{l l c}
	\midrule
    \#search results & doc. accuracy & sentence recall \\
    \midrule
    3    &   92.60    &   85.37 \\
	5    &   93.30    &   86.02 \\
    7    &   \textbf{93.55}    &   \textbf{86.24} \\
    \bottomrule
\end{tabular}
\caption{Performance of the retrieval systems using different numbers of MediaWiki search results}
\vspace{-1.5ex}
\label{tb:retrieval_search}
\end{table}

For the subtask of recognizing textual entailment, we also experiment with different numbers of selected sentences. The results in Table~\ref{tb:rte_sents} demonstrate that our model performs best with all five selected sentences.

\begin{table}[h]
\centering
\begin{tabular}{l l c }
  \toprule
  \#sentence(s) & label accuracy &   FEVER score \\ 
  \midrule
 1          & 60.03 	&  54.80   \\ 
 2          & 61.87 	&  57.19   \\
 3          & 64.29 	&  59.33   \\
 4          & 66.30 	&  61.79   \\
 5          & \textbf{68.49} 	&  \textbf{64.74}   \\
  \bottomrule
\end{tabular}
\caption{Performance of the textual entailment model using different numbers of sentences}
\vspace{-1.5ex}
\label{tb:rte_sents}
\end{table}

In Table~\ref{tb:results}, we compare the performance of our three systems as well as the full pipeline to the baseline systems and pipeline implemented 
by the shared task organizers \cite{Thorne18Fever} on the development set. 
As the results demonstrate, we were able to significantly improve upon the baseline on each sub-task.
The performance gains over the whole pipeline add up to an improvement of about 100\% 
with respect to the baseline pipeline.

\begin{table}[h]
\centering
\begin{tabular}{l l c }
  \toprule
  Task (metric) & system &   score (\%) \\
  \midrule
    Document retrieval   & baseline		& 70.20    \\ 
    (accuracy)           & our system   & 93.55    \\ 
  \midrule
  Sentence selection    & baseline      & 44.22    \\ 
      (recall)          & our system	& 87.10    \\ 
  \midrule
  Textual entailment    & baseline      & 52.09   \\ 
  (label accuracy)      & our system 	& 68.49   \\ 
  \bottomrule
  Full pipeline         & baseline      & \textbf{32.27}   \\ 
 (FEVER score)          & our system 	&  \textbf{64.74}   \\ 
  \bottomrule
\end{tabular}
\caption{Performance comparison of our system and the baseline system on the development set}
\vspace{-1.5ex}
\label{tb:results}
\end{table}

\section{Error analysis}

In this section, we present the error analysis for each of the three sub-tasks, which can serve as a basis for further improvements of the system.

\subsection{Document retrieval}
The typical errors encountered for the document retrieval system can be divided into three classes.

\noindent
\textbf{Spelling errors:} A word in the claim or in the article title is misspelled. E.g. 
``\emph{Homer Hickman wrote some historical fiction novels.}" vs. ``\emph{Homer Hickam}". In this case, our document retrieval system discards the article during the filtering phase.
 
\noindent
\textbf{Missing entity mentions:} The entity mention represented by the title of the article, which needs to be retrieved, is not related to any entity mention in the claim. E.g. Article title: ``\emph{Blue Jasmine}" Claim: ``\emph{Cate Blanchett ignored the offer to act in Cate Blanchett.}".

\noindent
\textbf{Search failures:} Some article titles contain a category name in parentheses for the disambiguation of the entity. This makes it difficult to retrieve the exact article title using the MediaWiki API. E.g. the claim ``\emph{Alex Jones is apolitical.}" requires the article ``\emph{Alex Jones (radio host)}", but it is not contained in the MediaWiki search results.
 
 \subsection{Sentence selection}

The most frequent case, in which the sentence selection model fails to retrieve a correct evidence set, is that the entity mention in the claim does not occur in the annotated evidence set. E.g. the only evidence set for the claim ``\emph{Daggering is nontraditional.}" consists of the single sentence ``\emph{This dance is not a traditional dance.}". 
Here, ``\emph{this dance}" refers to ``\emph{daggering}" and cannot be resolved by our model, since the information that ``\emph{daggering}" is a dance is not mentioned in the evidence sentence or in the claim.
Some evidence sets contain two sentences one of which is less related to the claim. E.g. the claim ``\emph{Herry II of France has three cars.}" has an evidence set that contains the two sentences ``\emph{Henry II died in 1559.}" and ``\emph{1886 is regarded as the birth year of the modern car.}". The second sentence is not directly related to the claim, thus, it is ranked very low by our model.


\subsection{Recognizing textual entailment}
A large number of claims are misclassified due to the model's disability to interpret numerical values.
For instance, the claim ``\emph{The heart beats at a resting rate close to 22 beats per minute.}" is not classified as \emph{refuted} on the basis of the evidence sentence ``\emph{The heart beats at a resting rate close to 72 beats per minute.}". The only information refuting the claim is the number, but neither GloVe nor FastText embeddings can embed numbers distinctly enough.
Another problem are challenging \emph{NotEnoughInfo} cases.
For instance, the claim ``\emph{Terry Crews played on the Los Angeles Chargers.}" (annotated as \emph{NotEnoughInfo}) is classified as \emph{refuted}, given the sentence ``\emph{In football, Crews played ... for the Los Angeles Rams, San Diego Chargers and Washington Redskins, ...}". The sentence is related to the claim but does not exclude it, which makes this case difficult.

\section{Conclusion}

In this paper, we presented the system for fact extraction and verification,
which we developed in the course of the FEVER shared task.
According to the preliminary results, our system scored third out of 23 competing teams.
The shared task was divided into three parts:
(i) Given a claim, retrieve Wikipedia documents that contain facts about the claim. 
(ii) Extract these facts from the document.
(iii) Verify the claim on the basis of the extracted facts.
To address the problem, we developed models for the three sub-tasks.
We framed document retrieval as entity linking by identifying entities 
in the claim and linking them to Wikipedia articles. 
To extract facts in the articles, 
we developed a sentence ranking model by extending the ESIM. 
For claim verification we proposed another extension to the ESIM, 
whereby we were able to classify the claim on the basis of multiple facts using attention.
Each of our three models, as well as the combined pipeline, significantly outperforms the baseline on the development set.

\section{Acknowledgements}

This work has been supported by the German Research Foundation as part of the Research
Training Group Adaptive Preparation of Information from Heterogeneous Sources (AIPHES) at the Technische Universit\"at Darmstadt grant No. GRK 1994/1. 

\bibliography{emnlp2018}
\bibliographystyle{acl_natbib_nourl}

\end{document}